\documentclass[aps,prl,showpacs,twocolumn,superscriptaddress,letterpaper,linenumbers]{revtex4} 

\usepackage{graphicx}
\usepackage{dcolumn}
\usepackage{bm}
\usepackage{amsmath, amssymb}%

\newcommand{\eepp}{\mbox{$(e,e'pp)$}}
\newcommand{\eep}{\mbox{($e,e'p$)}} 
\newcommand{\ee}{\mbox{$(e,e')$}}
\newcommand{\eeppn}{\mbox{$(e,e'pp)n$}}
 
\newcommand{\Het}{$^3$He}

\begin{document}


\title{A comparison of forward and backward $pp$ pair knockout in $^3$He$(e,e'pp)n$}

\newcommand*{\ODU}{Old Dominion University, Norfolk, Virginia 23529}
\newcommand*{\ODUindex}{26}
\affiliation{\ODU}
\newcommand*{\ANL}{Argonne National Laboratory, Argonne, Illinois 60439}
\newcommand*{\ANLindex}{1}
\affiliation{\ANL}
\newcommand*{\CANISIUS}{Canisius College, Buffalo, NY}
\newcommand*{\CANISIUSindex}{2}
\affiliation{\CANISIUS}
\newcommand*{\CMU}{Carnegie Mellon University, Pittsburgh, Pennsylvania 15213}
\newcommand*{\CMUindex}{3}
\affiliation{\CMU}
\newcommand*{\CUA}{Catholic University of America, Washington, D.C. 20064}
\newcommand*{\CUAindex}{4}
\affiliation{\CUA}
\newcommand*{\SACLAY}{CEA, Centre de Saclay, Irfu/Service de Physique Nucl\'eaire, 91191 Gif-sur-Yvette, France}
\newcommand*{\SACLAYindex}{5}
\affiliation{\SACLAY}
\newcommand*{\CNU}{Christopher Newport University, Newport News, Virginia 23606}
\newcommand*{\CNUindex}{6}
\affiliation{\CNU}
\newcommand*{\UCONN}{University of Connecticut, Storrs, Connecticut 06269}
\newcommand*{\UCONNindex}{7}
\affiliation{\UCONN}
\newcommand*{\FU}{Fairfield University, Fairfield CT 06824}
\newcommand*{\FUindex}{8}
\affiliation{\FU}
\newcommand*{\FIU}{Florida International University, Miami, Florida 33199}
\newcommand*{\FIUindex}{9}
\affiliation{\FIU}
\newcommand*{\FSU}{Florida State University, Tallahassee, Florida 32306}
\newcommand*{\FSUindex}{10}
\affiliation{\FSU}
\newcommand*{\Genova}{Universit$\grave{a}$ di Genova, 16146 Genova, Italy}
\newcommand*{\Genovaindex}{11}
\affiliation{\Genova}
\newcommand*{\GWUI}{The George Washington University, Washington, DC 20052}
\newcommand*{\GWUIindex}{12}
\affiliation{\GWUI}
\newcommand*{\ISU}{Idaho State University, Pocatello, Idaho 83209}
\newcommand*{\ISUindex}{13}
\affiliation{\ISU}
\newcommand*{\INFNFE}{INFN, Sezione di Ferrara, 44100 Ferrara, Italy}
\newcommand*{\INFNFEindex}{14}
\affiliation{\INFNFE}
\newcommand*{\INFNFR}{INFN, Laboratori Nazionali di Frascati, 00044 Frascati, Italy}
\newcommand*{\INFNFRindex}{15}
\affiliation{\INFNFR}
\newcommand*{\INFNGE}{INFN, Sezione di Genova, 16146 Genova, Italy}
\newcommand*{\INFNGEindex}{16}
\affiliation{\INFNGE}
\newcommand*{\INFNRO}{INFN, Sezione di Roma Tor Vergata, 00133 Rome, Italy}
\newcommand*{\INFNROindex}{17}
\affiliation{\INFNRO}
\newcommand*{\ORSAY}{Institut de Physique Nucl\'eaire ORSAY, Orsay, France}
\newcommand*{\ORSAYindex}{18}
\affiliation{\ORSAY}
\newcommand*{\ITEP}{Institute of Theoretical and Experimental Physics, Moscow, 117259, Russia}
\newcommand*{\ITEPindex}{19}
\affiliation{\ITEP}
\newcommand*{\JMU}{James Madison University, Harrisonburg, Virginia 22807}
\newcommand*{\JMUindex}{20}
\affiliation{\JMU}
\newcommand*{\KNU}{Kyungpook National University, Daegu 702-701, Republic of Korea}
\newcommand*{\KNUindex}{21}
\affiliation{\KNU}
\newcommand*{\LPSC}{LPSC, Universite Joseph Fourier, CNRS/IN2P3, INPG, Grenoble, France
}
\newcommand*{\LPSCindex}{22}
\affiliation{\LPSC}
\newcommand*{\UNH}{University of New Hampshire, Durham, New Hampshire 03824-3568}
\newcommand*{\UNHindex}{23}
\affiliation{\UNH}
\newcommand*{\NSU}{Norfolk State University, Norfolk, Virginia 23504}
\newcommand*{\NSUindex}{24}
\affiliation{\NSU}
\newcommand*{\OHIOU}{Ohio University, Athens, Ohio  45701}
\newcommand*{\OHIOUindex}{25}
\affiliation{\OHIOU}
\newcommand*{\RPI}{Rensselaer Polytechnic Institute, Troy, New York 12180-3590}
\newcommand*{\RPIindex}{27}
\affiliation{\RPI}
\newcommand*{\URICH}{University of Richmond, Richmond, Virginia 23173}
\newcommand*{\URICHindex}{28}
\affiliation{\URICH}
\newcommand*{\ROMAII}{Universita' di Roma Tor Vergata, 00133 Rome Italy}
\newcommand*{\ROMAIIindex}{29}
\affiliation{\ROMAII}
\newcommand*{\MSU}{Skobeltsyn Nuclear Physics Institute, Skobeltsyn Nuclear Physics Institute, 119899 Moscow, Russia}
\newcommand*{\MSUindex}{30}
\affiliation{\MSU}
\newcommand*{\SCAROLINA}{University of South Carolina, Columbia, South Carolina 29208}
\newcommand*{\SCAROLINAindex}{31}
\affiliation{\SCAROLINA}
\newcommand*{\JLAB}{Thomas Jefferson National Accelerator Facility, Newport News, Virginia 23606}
\newcommand*{\JLABindex}{32}
\affiliation{\JLAB}
\newcommand*{\UTFSM}{Universidad T\'{e}cnica Federico Santa Mar\'{i}a, Casilla 110-V Valpara\'{i}so, Chile}
\newcommand*{\UTFSMindex}{33}
\affiliation{\UTFSM}
\newcommand*{\GLASGOW}{University of Glasgow, Glasgow G12 8QQ, United Kingdom}
\newcommand*{\GLASGOWindex}{34}
\affiliation{\GLASGOW}
\newcommand*{\VT}{Virginia Polytechnic Institute and State University, Blacksburg, Virginia   24061-0435}
\newcommand*{\VTindex}{35}
\affiliation{\VT}
\newcommand*{\VIRGINIA}{University of Virginia, Charlottesville, Virginia 22901}
\newcommand*{\VIRGINIAindex}{36}
\affiliation{\VIRGINIA}
\newcommand*{\WM}{College of William and Mary, Williamsburg, Virginia 23187-8795}
\newcommand*{\WMindex}{37}
\affiliation{\WM}
\newcommand*{\YEREVAN}{Yerevan Physics Institute, 375036 Yerevan, Armenia}
\newcommand*{\YEREVANindex}{38}
\affiliation{\YEREVAN}

\newcommand*{\NOWVIRGINIA}{University of Virginia, Charlottesville, Virginia 22901}
\newcommand*{\NOWSACLAY}{CEA, Centre de Saclay, Irfu/Service de Physique Nucl\'eaire, 91191 Gif-sur-Yvette, France}
\newcommand*{\NOWMSU}{Skobeltsyn Nuclear Physics Institute, Skobeltsyn Nuclear Physics Institute, 119899 Moscow, Russia}
\newcommand*{\NOWINFNGE}{INFN, Sezione di Genova, 16146 Genova, Italy}
\newcommand*{\NOWANL}{Argonne National Laboratory, Argonne, Illinois 60439}

\author{H. Baghdasaryan}
     \altaffiliation[Current address: ]{\NOWVIRGINIA}
     \affiliation{\ODU}
\author{L.B.~Weinstein}
\email[Contact Author \ ]{weinstein@odu.edu}
     \affiliation{\ODU}
\author{J.M.~Laget}
\affiliation{\JLAB}
\author {K.P. ~Adhikari} 
\affiliation{\ODU}
\author {M.~Aghasyan} 
\affiliation{\INFNFR}
\author {M.J.~Amaryan} 
\affiliation{\ODU}
\author {M.~Anghinolfi} 
\affiliation{\INFNGE}
\author {J.~Ball} 
\affiliation{\SACLAY}
\author {M.~Battaglieri} 
\affiliation{\INFNGE}
\author {A.S.~Biselli} 
\affiliation{\FU}
\affiliation{\RPI}
% ...

\collaboration{The CLAS Collaboration}
\noaffiliation


\date{\today}

\begin{abstract}
 Measuring nucleon-nucleon Short Range Correlations (SRC) has been a
 goal of the nuclear physics community for many years.  They are an important part
  of the nuclear wavefunction, accounting for almost all of the
  high-momentum strength. They are closely related to the EMC effect.
While their overall probability has been measured, measuring their
momentum distributions is more difficult.   
In order to determine the best configuration for studying SRC
momentum distributions, we measured the $^3$He$(e,e'pp)n$ reaction, looking at events with high momentum
  protons ($p_p > 0.35$ GeV/c) and a low momentum neutron ($p_n< 0.2$ GeV/c).
  We examined two angular configurations: either both protons emitted forward
  or one proton emitted forward and one backward (with respect to the
  momentum transfer, $\vec q$).  The measured relative momentum
  distribution of the events with one forward and one backward proton
  was much closer to the calculated initial-state $pp$ relative
  momentum distribution, indicating that this is the preferred
  configuration for measuring SRC.
\end{abstract}


\maketitle

Early models of the nucleus described its structure in terms of
individual nucleons moving independently of each other in a
mean field.  However, this only describes about 70\% of the nucleus.
The missing 30\% is presumably due to nucleons in short and long
range correlations \cite{kelly96}.
  
Nucleon-nucleon ($NN$) Short Range Correlations (SRC) are a very important part of
the nuclear wave function.  The two nucleons in an SRC are at comparatively short
distances and thus higher densities than mean field nucleons.  These
SRC nucleons account for almost all of the high
momentum ($p>p_{fermi}\approx 0.25$ GeV/c) nucleons and most of the
kinetic energy in the nucleus.  Nucleons have a probability of
between $\approx 5$\%
(deuterium) and $\approx 25$\% ($A\ge 56$) of belonging to an SRC pair
\cite{egiyan02,egiyan06,fomin11}.  

SRC can
affect the rate of neutron star cooling
\cite{frankfurt08}.  In the direct Urca process ($p\rightarrow
n+e^++\nu$ and $n\rightarrow p + e^- + \bar\nu$), nucleons in the
neutron star beta-decay
and the neutrino carries energy away.
However, the decay nucleon is frequently at a momentum below the Fermi-surface and
then the process is Pauli-blocked.  In the modified
Urca process, a second nucleon is involved so that the decay products
are no longer below the Fermi-surface and the process is not
Pauli-blocked.  SRC, by moving nucleons from below to above the
Fermi-surface and opening holes in the Fermi-sphere, can also allow Urca-process cooling to occur.

In addition, it was recently found that the probability of a nucleon
belonging to an SRC in nucleus $A$ is remarkably closely correlated
with the strength of the EMC effect as measured in lepton Deep Inelastic
Scattering (DIS) in that nucleus
\cite{weinstein11}.  The EMC effect is the decrease in the per-nucleon
cross section of nucleus $A$ relative to deuterium.  This effect
cannot be explained without including some modification of the
nucleons in the nucleus \cite{kulagin10}.
The correlation indicates that SRC and the EMC effect stem
from the same underlying cause and that  nucleon
modification in the nucleus is related to SRC.

The relative probabilities of finding nucleons belonging to SRC have
been measured by inclusive $(e,e')$ experiments.  They measured the
per-nucleon cross section ratio of nucleus $A$ to
deuterium \cite{frankfurt93,fomin11} or $^3$He \cite{egiyan02,egiyan06}
at fixed four-momentum transfer $Q^2$ ($Q^2 = -q_\mu q^\mu = \vec
q\thinspace^2 - \nu^2$, $\nu$ is the energy transfer, and $\vec
q$ is the three-momentum transfer) as a function of
$x=Q^2/2m\nu$ where $m$ is the nucleon mass.  There
is a plateau in the cross section ratio for $Q^2 > 1.5$ GeV$^2$
ranging from about $1.5 < x < 2$.   Under certain reasonable
assumptions (see references for details), the minimum initial
struck-nucleon momentum is a function of only $x$ and $Q^2$.  The existence
of this plateau indicates that nucleons have
the same momentum distribution in different nuclei for momenta greater
than some threshold.  The location of the onset of the
plateau in $x$ and $Q^2$ indicates that this threshold is $p_{thresh}= 0.275\pm0.025$
GeV/c.  The height of the plateau (the magnitude of the per-nucleon
cross section ratio) corresponds to the relative probability of
finding nucleons in the two nuclei for $p>0.275$ GeV/c.

Since the different nuclei ($2\le A\le 197$)
have very different characteristics (density, radius, etc.), this
similar momentum distribution at high momentum cannot be due to the
$A-1$ other nucleons and thus can only be due to the presence of a
single adjacent nucleon, {\it i.e.,} due to $NN$
SRC.  Thus the value of the per-nucleon cross section ratio in the
plateau region equals the
relative probability that nucleons in the two nuclei belong to short
range correlations.

It is more difficult to measure the relative and total momentum
distributions of the correlated nucleons.  Measurements of the
$^3$He$(e,e'pp)n$ reaction studied events where the virtual photon is absorbed
by the third nucleon and the other two nucleons belong to a
spectator correlated pair which disintegrates in the absence of the
third nucleon \cite{niyazov03,bagh10}.  The relative and total final
state momenta, $p_{rel}$ and $p_{tot}$, of the other two nucleons then
should correspond to those quantities in the initial state.  This
measurement is complicated by the strong continuum-state interaction
between those two nucleons in the final state.  This technique also
does not apply to nuclei heavier than $^3$He.

Efforts to measure the momentum distributions of correlated $NN$ pairs
in heavier nuclei focus on knocking out a high-initial-momentum
nucleon (usually  a proton) and detecting its correlated partner.
This can be done with either proton \cite{tang03,piasetzky06} or
electron \cite{shneor07,subedi08} probes.  However, the interpretation
of these experiments 
can be complicated by the final state interaction of the knocked-out proton,
as well as from the effects of two-body currents such as meson exchange
currents and isobar configurations ({\it e.g.,} $\Delta(1232)$ production followed
by de-excitation and absorption of the decay pion on another nucleon).

There are two general kinematic configurations for these measurements.  The partner
nucleon can be detected in the forward hemisphere relative to the
momentum transfer $\vec q$ or in the backward hemisphere.  If the
partner nucleon is detected in the forward hemisphere, then the
magnitude of the momentum of the struck nucleon will be {\it less
  than} the magnitude of the three-momentum transfer, $\vert \vec p_s
\thinspace\vert < \vert \vec q\thinspace\vert$.  Compared to a nucleon
of final momentum $\vec q$, the
kinetic energy of the struck nucleon will be smaller and the
energy transfer, $\nu$, will also be smaller so that $x>1$.  If the
partner nucleon is detected in the backward hemisphere, then the
magnitude of the momentum of the struck nucleon will be {\it greater
  than} the magnitude of the three-momentum transfer, $\vert \vec p_s
\thinspace\vert > \vert \vec q\thinspace\vert$.  In this case, the
kinetic energy of the struck nucleon and hence the energy transfer
will be greater so that $x<1$.

Ref.~\cite{shneor07} argues that the forward kinematics with
smaller $\nu$ and $x>1$ is preferred.  They argue that these
conditions, on the low-energy-transfer side of the $(e,e')$ quasielastic peak,
are farther from the region where meson exchange currents and
$\Delta$-production can contribute.  Therefore, cross sections
measured at forward kinematics should be more sensitive to the short-range
nucleon-nucleon correlations, {\it i.e.,} to the initial state
momentum distribution.

On the contrary, Ref.~\cite{meln97} argues that the configuration
where one proton is emitted backward is preferred.  They argue that it
is very difficult for Final State Interactions (FSI) to produce backward
nucleons and therefore cross sections measured at backward kinematics
will be more sensitive to the nuclear initial state.  This argument is
supported by $d(e,e'p)$ measurements \cite{klim06} where cross
sections measured at backward proton angles agreed well
with calculations that did not include FSI.

The present paper reports measurements of two proton knockout from $^3$He in both
forward, $x>1$, and backward, $x<1$, kinematics in order to compare the
measured relative momentum distributions and to determine which
kinematic configuration is more sensitive to the initial state
momentum distribution.

We measured the \Het\eeppn{} reaction at the Thomas Jefferson National Accelerator
Facility (Jefferson Lab) in 2002 using a 100\% duty factor, 5--10 nA beam of
4.7 GeV electrons incident on a 5-cm liquid $^3$He or H$_2$ target.  We
detected the outgoing charged particles in the
CEBAF Large Acceptance Spectrometer (CLAS) \cite{clas}.  

CLAS uses a toroidal magnetic field (with in-bending electrons) and
six independent sets of drift chambers, time-of-flight scintillation
counters and electro-magnetic calorimeters (EC) for charged particle
identification and trajectory reconstruction.  The polar angular
acceptance is $8^\circ < \theta < 140^\circ$ and the azimuthal angular
acceptance is 50\% at smaller polar angles, increasing to 80\% at
larger polar angles.  The EC was used for the electron trigger with a
threshold of approximately 0.9 GeV.  

We eliminated the effects of interactions in the target walls by requiring particles
to come from the central 4-cm of the target.  We identified electrons
using the energy deposited in the EC, and protons using
time-of-flight.  The H\eep{} cross section was measured
and compared to a parametrization of the world's cross section data
\cite{arrington03} to determine our electron and proton detection
efficiencies \cite{HBPhD}.

Regions of non-uniform detector
response were excluded by software cuts, while acceptance and tracking
efficiencies were estimated using GSIM, the CLAS GEANT Monte-Carlo
simulation \cite{gsim}.  Momentum coverage extended down to 0.35 GeV/c for
protons.

\begin{figure}[htbp]
    \includegraphics[height=2.3in]{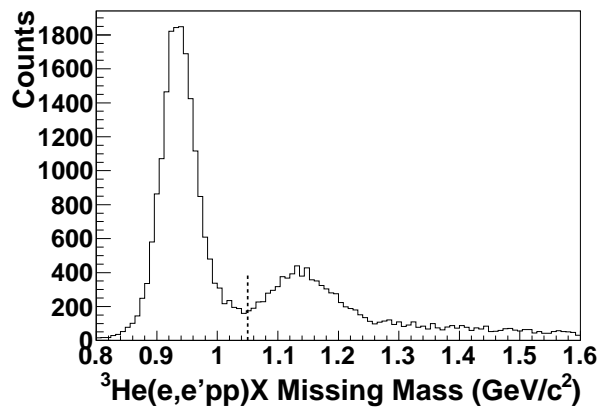} \caption{
      Missing mass for \Het\eepp$X$ for missing momentum $p_X < 0.2$ GeV/c.  The dashed
      vertical line
    indicates the neutron missing mass cut of $M_X<1.05$ GeV/c$^2$. 
       \label{fig:missmass} }
\end{figure}

\begin{figure}[htbp]
  \includegraphics[height=2.3in]{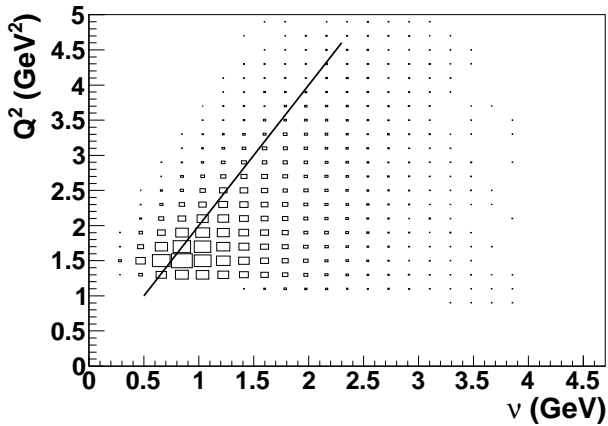} \caption{
    The square of the four-momentum transfer, $Q^2$, versus the energy
    transfer $\nu$ for \Het\eeppn{} events with $p_n\le 0.2$ GeV/c.  The points show the data, the straight
    line shows quasielastic kinematics where $x=Q^2/2m\nu=1$.  The
    lower limit at $Q^2\approx 1$ GeV$^2$ is due to the CLAS acceptance.
     \label{fig:Q2omegaUncut} }
\end{figure}

We identified the neutron using a
missing mass cut to select \Het\eeppn{} events (see Fig.~\ref{fig:missmass}).  We required that each
neutron had momentum $p_n \le 0.2$ GeV/c in order to focus on $pp$
pairs with small total momentum.  (Events with $p_n \ge 0.25$ GeV/c are discussed in Ref.~\cite{bagh10}.)
Fig.~\ref{fig:Q2omegaUncut} shows that the experiment covered a wide
range of energy and momentum transfers.  For \Het\eeppn{} events, the
momentum transfer $Q^2$ peaks at around 1.5 GeV$^2$.  The energy
transfer $\nu$ is
concentrated slightly above but close to quasielastic kinematics
($\nu \approx Q^2 / 2m_p$ or $x \approx 1$).

\begin{figure}[htbp]
    \includegraphics[height=2.3in]{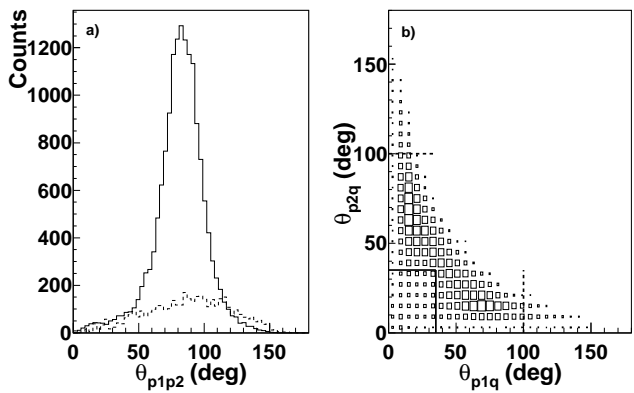} \caption{a) The
      number of counts as a function of opening angle between the two protons in the lab frame for
      \Het\eeppn{} events for $p_n < 0.2$ GeV/c for data (solid
      histogram) and the one-body calculation of Laget integrated over
      the experimental acceptance with arbitrary
      normalization (dashed histogram).  b) The number of
      counts as a function of the two
      proton-momentum transfer angles, $\theta_{p_1q}$ and
      $\theta_{p_2q}$.
      \label{fig:ppopen} }
\end{figure}

Since the two protons shared the energy and momentum transfer of the
reaction, we looked at the opening angle of the two protons (see
Fig.~\ref{fig:ppopen}a).  The distribution peaks at an opening
angle of about 80$^\circ$, characteristic of final state rescattering.
(Nonrelativistically and classically, if one proton hits a second proton at rest, then
the opening angle in the final state will be exactly 90$^\circ$.)  A
one-body cross section
calculation by Laget (described in detail below) integrated over the
experimental acceptance does not show this rescattering peak,
indicating that it is not an artifact of the experimental acceptance.

A two-dimensional plot
of the opening angle between each proton and the momentum transfer,
$\vec q$, (see Fig.~\ref{fig:ppopen}b) shows peaks where one proton is
at an angle of 70$^\circ$ with respect to $\vec q$ and the other proton is at
about $15-20^\circ$.  These peaks are indicative of small angle
rescattering, where one proton absorbs the virtual photon and scatters
from the second proton in the final state.  The first proton is
slightly deflected from its original direction and the second proton
is scattered at about $70^\circ$.

\begin{figure}[htbp]
    \includegraphics[height=3.2in]{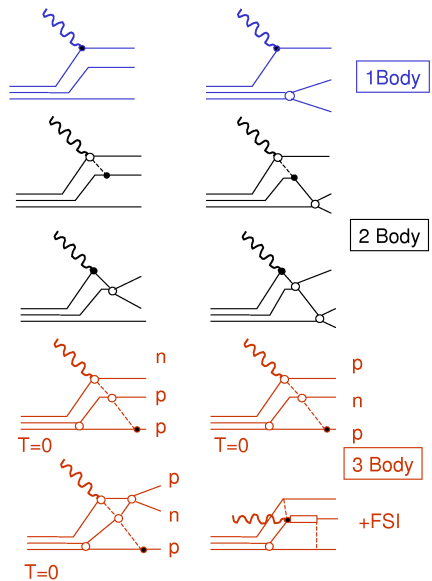} 
    \caption{(color online) The relevant graphs in the diagrammatic calculation of
      the \Het\eeppn{} reaction \cite{laget87,laget88,laget94,laget05} including
      one-body, two-body, and three body
      mechanisms.  The graph corresponding to Final State Interactions
      (as the term is used in this paper)
      is the middle graph on the left side.
     \label{fig:diagrams} }
\end{figure}

In order to study the contribution of different
reaction mechanisms in different experimental configurations, we compared
our data with the diagrammatic calculation of Laget.  This calculates the differential cross section from the square of the
coherent sum of the amplitudes associated with the diagrams in
Fig.~\ref{fig:diagrams}: the one-body, the two-body Final State Interactions
 or Meson Exchange Currents (MEC) and the three-body
mechanisms. The antisymmetric bound state wave function is the
solution~\cite{hajduk79} of the Faddeev equations for the Paris
potential~\cite{lac81}. The continuum is approximated by the combination of the
plane wave amplitudes and half off shell amplitudes where two nucleons scatter,
the third being spectator. The antisymmetry of the final state is
achieved by interchanging the role of the three nucleons. Two body MEC
are computed as described in~\cite{laget94}, while three body mechanisms
are implemented as in~\cite{laget88}.

Note that the one-body cross section is
proportional to the initial state momentum distribution.  This will be
useful in helping us identify kinematic regions where the cross
section is sensitive to the initial state.

The theoretical cross sections are then integrated by a Monte Carlo
sampling of the phase space within the fiducial acceptance of
CLAS and then binned in the
same way as the experimental data.

At low energy, the application of the model to our channel, the
electrodisintegration of a $pp$ pair at rest, has been described in
\cite{laget87}. It has been adapted to higher energy according to
\cite{laget05}. The nucleon $s$-wave scattering has been supplemented by
a high energy diffractive scattering amplitude that fits the
experimental $NN$ cross section. It uses a fully relativistic nucleon
current with the latest experimental values of the nucleon form
factors. It describes well the two body
\cite{rvachev05} and three body \cite{benmokhtar05} break up of \Het{}
recently studied at Jefferson Lab in the same energy and momentum range.

The rescattering peak, near 70$^\circ$ in Fig.~\ref{fig:ppopen}b, is more prominent
than in the $^2$H$(e,e'p)n$ reaction \cite{egi07} under similar
kinematics. The reason is that a $pp$ pair at rest is almost entirely in a
relative $s$-wave which has a node around 400 MeV/c: consequently the
one-body contribution is strongly suppressed. Also, unlike in the $pn$
channel~\cite{egi07}, the contribution of the $\Delta N$ intermediate
state to the $pp$ channel is very small. The reason is that a $pp$
pair has no dipole moment for the virtual photon to couple to. If a
virtual photon is absorbed on a $pp$ pair at rest, it would create a
$\Delta N$ system in a $1^+$ state which cannot then decay into a
$pp$ system (see e.g., Ref. \cite{laget87}).

\begin{figure}[htbp]
  \includegraphics[height=2.3in]{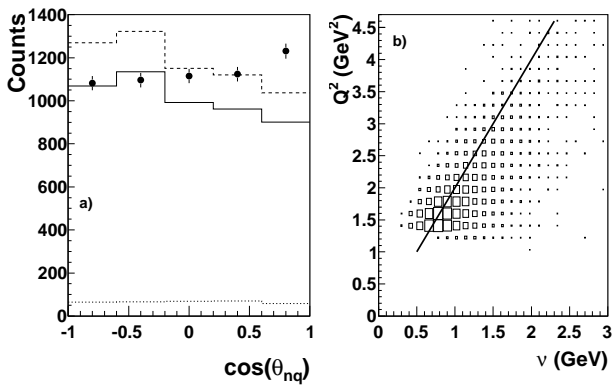} \caption{
    \Het\eeppn{} events with $p_n < 0.2$ GeV/c and $0.4\le
    p_p^{slow}\le 0.6$ GeV/c (where $p_p^{slow}$ is the smaller of the two
    proton momenta): a) The number of counts plotted versus
    $\cos\theta_{nq}$, the angle between the neutron momentum and
    $\vec q$.  The points show the data, the dotted
    curve shows Laget's one-body calculation, the dashed curve
    includes final state interactions, and the solid curve shows the
    full calculation including FSI and meson exchange currents
    \cite{laget87,laget88,laget94,laget05}. For ease of comparison of
    the angular distributions, the theoretical calculations are all multiplied by the
    same arbitrary factor to approximately scale the full calculation
    to the data.  The data and the calculations are all
    approximately isotropic.   b) The number of counts
    plotted as a function of the four-momentum transfer squared and
    the energy transfer ($Q^2$ and $\nu$).  The points show the
    data, the straight line shows quasielastic kinematics where
    $x=Q^2/2m\nu=1$.
    \label{fig:pslowQ2Theta} }
\end{figure}

In order to test the hypothesis that most of the $^3$He$(e,e'pp)n$ events are dominated by
final state rescattering and to validate the calculation, we looked at
the distribution of events with 0.4
to 0.6 GeV/c protons. These protons will always be the slower of the two protons in the
reaction.  This momentum range was selected to
maximize the expected effects of final state interactions.  

If these events are dominated by proton knockout followed by $pp$
rescattering, then the neutron should be spectator to the reaction and
its angular distribution with respect to
the momentum transfer $\vec q$ should be isotropic (see
Fig.~\ref{fig:pslowQ2Theta}a).  This agrees with the calculation both
without and with FSI.  The inclusion of FSI increases the
magnitude of the calculation by more than an order of magnitude and
the inclusion of Meson Exchange Currents
(MEC) changes it by only another 20\%.

The momentum and energy transfer distribution for these events is shown in
Fig.~\ref{fig:pslowQ2Theta}b.  The energy and momentum
transfers are centered around $x=1$, as expected for quasielastic
knockout with or without subsequent rescattering.

\begin{figure}[htbp]
  \includegraphics[height=2.8in]{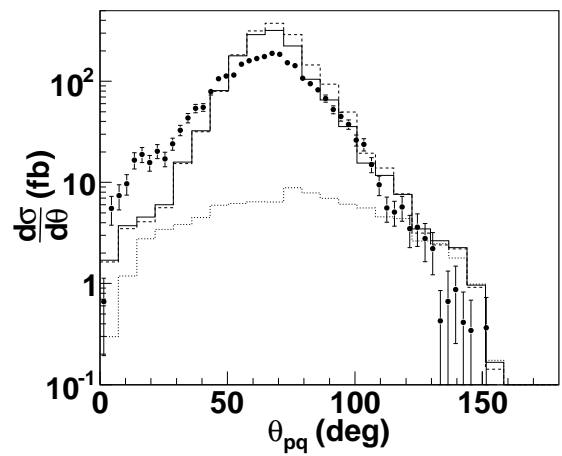} \caption{
    The laboratory-frame cross section for \Het\eeppn{} events
    with $p_n < 0.2$ GeV/c and $0.4\le p_p^{slow}\le 0.6$ plotted versus $\theta_{pq}$, the
    opening angle between the  proton momentum, $\vec p_p^{\thinspace slow}$,  and the momentum transfer $\vec q$.
 The points show the data, the
    dotted curve shows Laget's one-body calculation, the dashed curve
    includes final state interactions, and the solid curve shows the
    full calculation including FSI and meson exchange currents
    \cite{laget87,laget88,laget94,laget05}.   No scale factors have
    been applied to the calculations.  Systematic uncertainties of
    15\% are not shown.
          \label{fig:thetapslow} }
\end{figure}

Fig.~\ref{fig:thetapslow} shows the cross section as a function of the
angle between the slower proton ($0.4 \le p_p^{slow}\le 0.6$ GeV/c) and $\vec
q$.  The cross sections are corrected for radiative effects and
tracking efficiency and then integrated over the experimental
acceptance \cite{HBPhD}.  The systematic uncertainty is 15\%,
primarily due to the uncertainty in the low momentum proton detection
efficiency.  The data distribution has a prominent peak at around
$70^\circ$.  The calculation including the effects of FSI agrees
qualitatively with the data, peaking at around $\theta_{pq}\approx
70^\circ$ at a cross section more than ten times larger than that of
the one-body calculation.  This shows that the prominent peaks seen at
$\theta_{pq}\approx 70^\circ$ in Fig.~\ref{fig:ppopen}b are due to
FSI.  Note that the model only takes into account the dominant
diffractive scalar part of the $pp$ scattering amplitude, and the
inclusion of the full experimental amplitude (from the SAID data base
for instance) may likely improve the agreement between data and model.

The comparison between data and calculation for $0.4 \le p_p^{slow}\le 0.6$
GeV/c shows that the cross section is dominated by FSI and that the
calculation qualitatively agrees with the data.

The next step in the analysis was to try to identify kinematic configurations that are
sensitive to high-momentum components of the momentum distribution and
which are not significantly affected by FSI.  (MEC are suppressed for $pp$ knockout because
the virtual photon does not couple strongly to neutral pions.)  To do that we looked at events where
one proton is emitted backward with respect to the momentum transfer,
$\theta_{pq}\ge 100^\circ$, as shown by the dashed lines in
Fig.~\ref{fig:ppopen}b.  Because it is difficult for FSI to produce
backward-going nucleons, many theorists expect that this kinematics
will be the most sensitive to the nuclear initial state
\cite{meln97}.  

We also looked at events where both protons are emitted forward.
This kinematics was chosen because it corresponds to $x>1$, on the low
energy-loss side of the quasielastic peak, where there are smaller
contributions to the \ee{} cross section from meson exchange currents
and delta isobar currents.  Several short range correlations experiments \cite{shneor07,src07} 
have chosen to measure cross sections at 
$x>1$.  However, when both protons are
emitted forward they have smaller relative momentum in the final
state and hence will have larger final state interactions.  This
increased FSI will complicate the interpretation of the data.

\begin{figure}[htbp]
  \includegraphics[height=1.1in]{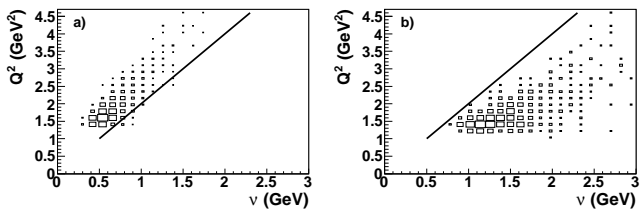} \caption{
    The square of the four-momentum transfer, $Q^2$, versus the energy
    transfer $\nu$ for \Het\eeppn{} events with $p_n\le 0.2$ GeV/c;
a)
    with two forward protons.  Both protons are within 35$^\circ$ of the
    momentum transfer, $\theta_{p_1q}$ and $\theta_{p_2q} \le 35^\circ$;
    and b) with one forward and one backward proton such that
    $\theta_{pq}\ge 100^\circ$.  The points show the data, the straight
    line shows quasielastic kinematics where $x=Q^2/2m\nu=1$.
     \label{fig:Q2omegaFwdBack} }
\end{figure}

\begin{figure}[htbp]
    \includegraphics[height=2.3in]{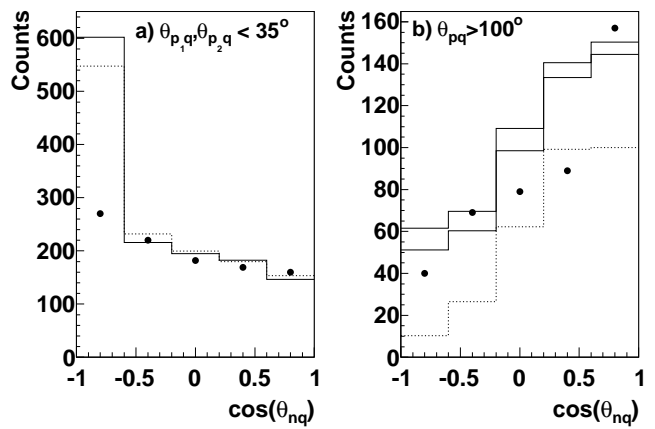} \caption{
      The angular distribution of $p_n \le 0.2$ GeV/c neutrons with respect to the
      momentum transfer $\vec q$, for \Het\eeppn{} events a) with two
      forward protons, $\theta_{p_1 q},\theta_{p_2 q}\le 35^\circ$, and b)
      with one backward proton $\theta_{pq}\ge 100^\circ$.
      The points show the data, the histograms are the same as in
      Fig. \ref{fig:thetapslow}.  For ease of comparison of
    the angular distributions, the calculations have been
      separately normalized for the two plots so that the full
      calculation (solid line) is approximately equal to the data.
     \label{fig:thetan2} }
\end{figure}

\begin{figure}[htbp]
  \includegraphics[height=2.3in]{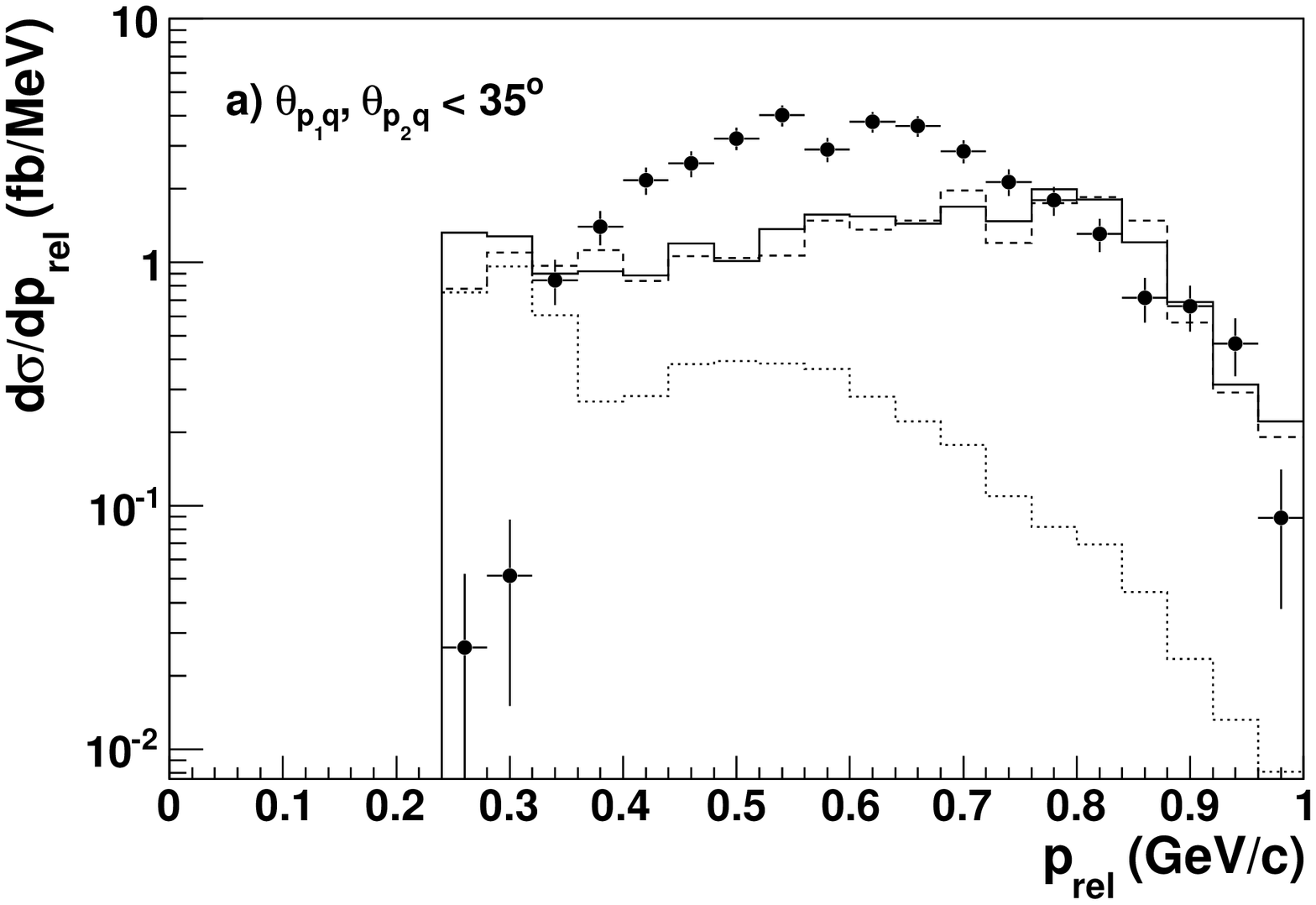}
   \includegraphics[height=2.3in]{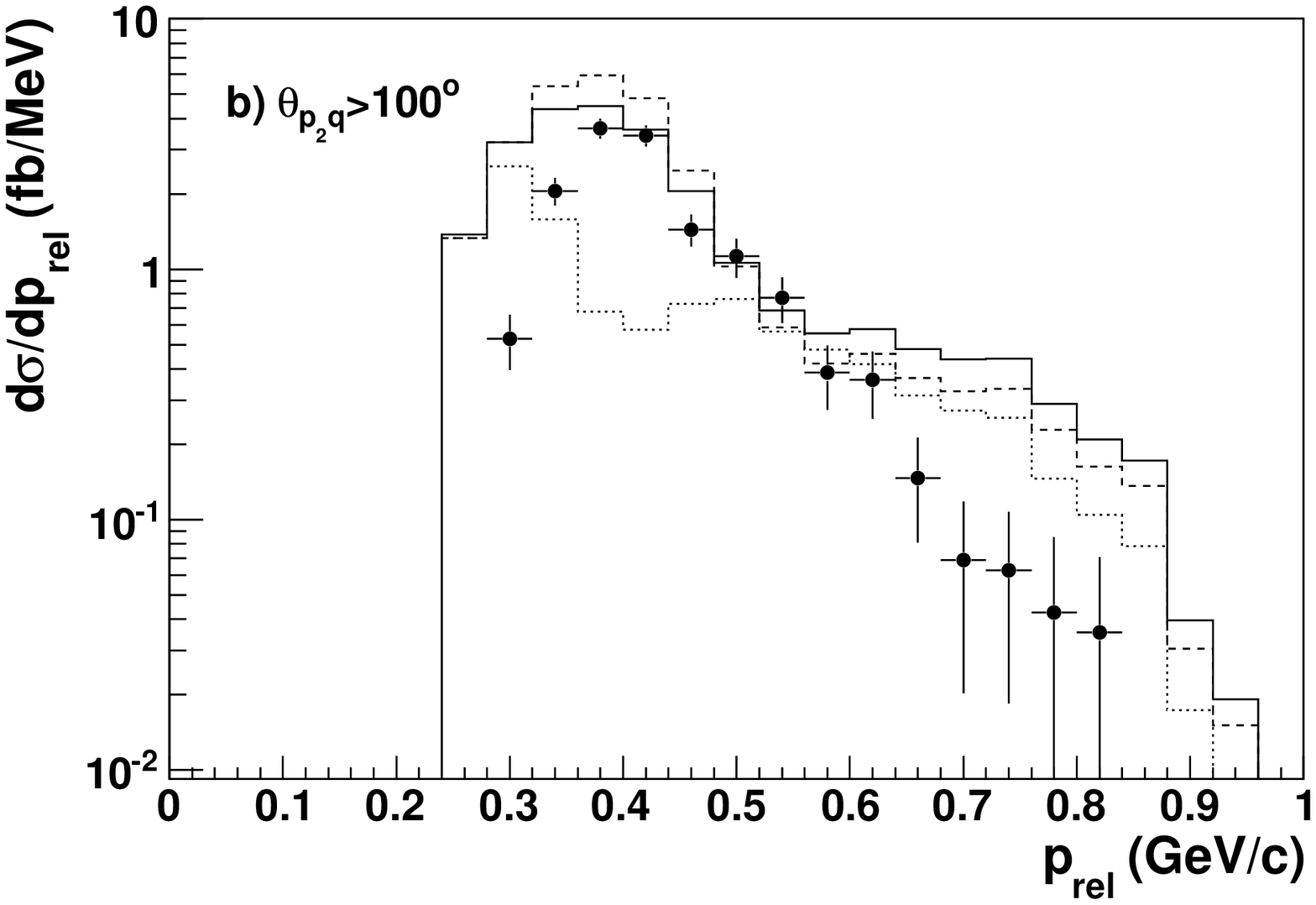}
      \caption{The \Het\eeppn{} laboratory-frame cross section plotted versus the
        ``$q$-subtracted'' relative momentum,
        $p_{rel}=\vert \vec p_1 - \vec q - \vec p_2
        \thinspace\vert/2$, for  events with a) both protons forward,
        $\theta_{p_1 q}, \theta_{p_2 q} \le 35^\circ$ and b)  one
        proton backward, $\theta_{p_2 q} \ge 100^\circ$.  The points show
        the data, the dotted line shows the one-body calculation, the
        dashed line shows the one-body + FSI calculation, and the solid
        line shows the full calculation including FSI and MEC. No scale factors have been applied to the calculations.
     \label{fig:prel} }
\end{figure}

Fig.~\ref{fig:Q2omegaFwdBack} shows the momentum- and energy-transfer distributions
for the two-forward-proton (``forward'') and the backward-proton
(``backward'') kinematics.  As expected, the forward kinematics is at lower
energy loss and $x>1$ and the backward kinematics is at higher
energy loss and $x<1$.  The neutron angular distributions for the two
kinematics are shown  in Fig.~\ref{fig:thetan2}.  These
neutron distributions are not as isotropic as for the events with moderate
momentum protons (see Fig.~\ref{fig:pslowQ2Theta}b).  This is due to the
initial state momentum distribution, since it is also seen in the
angular distribution of Laget's one-body
calculation.

For both forward and backward kinematics we calculated the
``$q$-subtracted'' relative momentum, 
\[
p_{rel}=\frac12 \vert \vec p_1 - \vec q - \vec p_2
        \thinspace\vert \quad .
\]
For the forward kinematics, we chose proton 1 such that $p_1 > p_2$.
For the backward kinematics we chose proton 1 to be the forward
proton.  In the one-body limit ({\it i.e.,} in the absence of FSI, MEC and
IC), the ``$q$-subtracted'' relative momentum equals the relative momentum of the two protons in the
initial state.  

The measured relative momentum distributions for the forward and
backward kinematics (see Fig.~\ref{fig:prel}) are very different.  The
forward proton $p_{rel}$ distribution peaks at significantly higher
momentum and is much broader than the backward proton distribution.
This strongly indicates that at least one of the two distributions is
not sensitive to the initial state momentum distribution.  
The minimum detected proton momentum of $p_p > 0.35$ GeV/c restricts
the minimum measurable relative momentum.

Calculations by Laget indicate that the forward proton kinematics is
much more sensitive to FSI.  FSI only affects the backward kinematics
at $p_{rel}\approx 0.4$ GeV/c by filling in the minimum of the $pp$
momentum distribution.  However, FSI increases the cross section by a
factor of at least three in
forward kinematics for all relative momenta greater than 0.4 GeV/c.
This is not surprising, since the final state relative momentum,
$p^f_{rel}=\vert \vec p_1 - \vec p_2 \vert/2$, is
significantly smaller for the forward kinematics, leading to a larger
final state interaction.

The Laget calculation agrees qualitatively with the data for
$p_{rel}>0.4$ GeV/c in both kinematics.  In the forward kinematics the
full cross section is several times larger than the one-body cross
section at $p_{rel} > 0.4$ GeV/c, indicating that the cross section is
dominated by the effects of FSI (see
Fig.~\ref{fig:prel}a).  Note that $pp$ rescattering in the final state (FSI)
redistributes strength from lower to higher relative momenta.  Thus,
the calculated cross section at large relative momenta depends on the
initial state momentum distribution at much smaller momenta and on the
details of the $pp$ rescattering model.

The full calculation underestimates the forward proton cross section for
$0.4 < p_{rel}<0.7$ GeV/c.  Because the cross section in this region
is dominated by FSI, this underestimate indicates the need to include
the full rescattering amplitude (including spin-dependent parts) and
not just its dominant diffractive (scalar) part.

In the
backward kinematics (Fig.~\ref{fig:prel}b) Laget's calculation overestimates
the backward proton cross sections for $p_{rel}>0.6$ GeV/c.  The full
calculation is very close to the one-body calculation in this region,
indicating that FSI effects are small and that the cross section is dominated by the initial state
momentum distribution.  Thus this overestimate indicates that the wave
function used (a Faddeev solution of the Paris potential) contains too
much high-momentum strength.  Disagreement at these large momenta is
not surprising, because $NN$ potential models (and hence calculated initial
state momentum distributions) are poorly constrained by elastic
scattering data for $p_{rel} > 0.35$ GeV/c, where inelastic channels
open up.  

Using a different nuclear wave function with a different initial state
momentum distribution could increase agreement between data and
calculation at the backward kinematics without decreasing agreement at
the forward kinematics.  The forward kinematics cross section is
dominated by FSI, that is by $pp$ rescattering which redistributes
strength from lower to higher relative momenta.  Therefore decreasing
the strength of the initial state momentum distribution at
$p_{rel}>0.6$ GeV/c would not significantly change the calculated
cross sections at forward kinematics and therefore would not affect
the agreement between data and calculation for the forward kinematics
at $p_{rel}>0.6$ GeV/c.


In summary, the large kinematics coverage of 
CLAS allowed us to identify the important reaction ingredients in various portions of 
the phase space when the neutron is almost a spectator. The data confirm 
the expectations of a model which combines the dominant parts of the 
amplitudes: the Faddeev three body wave function, final state
interactions using the dominant 
diffractive scalar part of the $NN$ scattering amplitude, and the MEC and 
$\Delta$ formation amplitudes. Proton-proton rescattering dominates the cross section 
around $x=1$. MEC and $\Delta$ formation contributions are relatively
small. 

We compared the $q$-subtracted relative momentum distributions for
kinematics with a low-momentum (spectator) neutron in two configurations:
(1) with both protons emitted in the forward direction and (2) with
one proton emitted forward and the other backward.   The full
calculation and the data agree qualitatively at both kinematic
configurations.   The calculation
shows that for $p_{rel}>0.5$ GeV/c, FSI are very small at the backward
kinematics but are dominant at the forward kinematics.
Thus, the cross section measured at backward kinematics is much closer
than that measured at forward
kinematics to the one-body calculation,
and hence is much more sensitive to the initial state momentum
distribution.

This result indicates that short range correlations studies using two-nucleon
knockout experiments to measure the $NN$ relative momentum
distribution should concentrate on kinematics at $x<1$ where one of the
nucleons is emitted backward.

\begin{acknowledgments}
We  acknowledge the outstanding efforts of the staff of
the Accelerator and Physics Divisions (especially the CLAS target
group) at Jefferson Lab that made this experiment possible.
This work was supported in part by the Italian Istituto Nazionale di Fisica
Nucleare, the Chilean 
CONICYT, the French Centre National de la Recherche Scientifique and
Commissariat \`{a} l'Energie Atomique, the UK Science and Technology Facilities Council (STFC), the U.S. Department of
Energy and National Science Foundation, 
and the National Research Foundation of Korea.  Jefferson Science Associates, LLC,
 operates the Thomas Jefferson National Accelerator
Facility for the United States Department of Energy under contract
DE-AC05-060R23177.

\end{acknowledgments}


\begin{thebibliography}{30}
\expandafter\ifx\csname natexlab\endcsname\relax\def\natexlab#1{#1}\fi
\expandafter\ifx\csname bibnamefont\endcsname\relax
  \def\bibnamefont#1{#1}\fi
\expandafter\ifx\csname bibfnamefont\endcsname\relax
  \def\bibfnamefont#1{#1}\fi
\expandafter\ifx\csname citenamefont\endcsname\relax
  \def\citenamefont#1{#1}\fi
\expandafter\ifx\csname url\endcsname\relax
  \def\url#1{\texttt{#1}}\fi
\expandafter\ifx\csname urlprefix\endcsname\relax\def\urlprefix{URL }\fi
\providecommand{\bibinfo}[2]{#2}
\providecommand{\eprint}[2][]{\url{#2}}

\bibitem[{\citenamefont{Kelly}(1996)}]{kelly96}
\bibinfo{author}{\bibfnamefont{J.}~\bibnamefont{Kelly}}, \bibinfo{journal}{Adv.
  Nucl. Phys.} \textbf{\bibinfo{volume}{23}}, \bibinfo{pages}{75}
  (\bibinfo{year}{1996}).


\bibitem[{\citenamefont{Egiyan et~al.}(2003)}]{egiyan02}
\bibinfo{author}{\bibfnamefont{K.}~\bibnamefont{Egiyan}} \bibnamefont{et~al.}
  (\bibinfo{collaboration}{CLAS Collaboration}), \bibinfo{journal}{Phys. Rev.
  C} \textbf{\bibinfo{volume}{68}}, \bibinfo{pages}{014313}
  (\bibinfo{year}{2003}).

\bibitem[{\citenamefont{Egiyan et~al.}(2006)}]{egiyan06}
\bibinfo{author}{\bibfnamefont{K.}~\bibnamefont{Egiyan}} \bibnamefont{et~al.}
  (\bibinfo{collaboration}{CLAS Collaboration}), \bibinfo{journal}{Phys. Rev.
  Lett.} \textbf{\bibinfo{volume}{96}}, \bibinfo{pages}{082501}
  (\bibinfo{year}{2006}).

\bibitem[{\citenamefont{Fomin et~al.}(2012)}]{fomin11}
\bibinfo{author}{\bibfnamefont{N.}~\bibnamefont{Fomin}} \bibnamefont{et~al.},
  \bibinfo{journal}{Phys. Rev. Lett.} \textbf{\bibinfo{volume}{in press}}
  (\bibinfo{year}{2012}).

\bibitem[{\citenamefont{Frankfurt and Strikman}(2008)}]{frankfurt08}
\bibinfo{author}{\bibfnamefont{L.}~\bibnamefont{Frankfurt}} \bibnamefont{and}
  \bibinfo{author}{\bibfnamefont{M.}~\bibnamefont{Strikman}}
  (\bibinfo{publisher}{AIP}, \bibinfo{year}{2008}),
  \bibinfo{editor}{\bibnamefont{eds. S.~Boffi, et al.}}, vol.
  \bibinfo{volume}{1056}, pp. \bibinfo{pages}{241--247}.

\bibitem[{\citenamefont{Weinstein et~al.}(2011)\citenamefont{Weinstein,
  Piasetzky, Higinbotham, Gomez, Hen, and Shneor}}]{weinstein11}
\bibinfo{author}{\bibfnamefont{L.~B.} \bibnamefont{Weinstein}},
  \bibinfo{author}{\bibfnamefont{E.}~\bibnamefont{Piasetzky}},
  \bibinfo{author}{\bibfnamefont{D.~W.} \bibnamefont{Higinbotham}},
  \bibinfo{author}{\bibfnamefont{J.}~\bibnamefont{Gomez}},
  \bibinfo{author}{\bibfnamefont{O.}~\bibnamefont{Hen}}, \bibnamefont{and}
  \bibinfo{author}{\bibfnamefont{R.}~\bibnamefont{Shneor}},
  \bibinfo{journal}{Phys. Rev. Lett.} \textbf{\bibinfo{volume}{106}},
  \bibinfo{pages}{052301} (\bibinfo{year}{2011}).

\bibitem[{\citenamefont{Kulagin and Petti}(2010)}]{kulagin10}
\bibinfo{author}{\bibfnamefont{S.~A.} \bibnamefont{Kulagin}} \bibnamefont{and}
  \bibinfo{author}{\bibfnamefont{R.}~\bibnamefont{Petti}},
  \bibinfo{journal}{Phys. Rev. C} \textbf{\bibinfo{volume}{82}},
  \bibinfo{pages}{054614} (\bibinfo{year}{2010}).

\bibitem[{\citenamefont{Frankfurt et~al.}(1993)\citenamefont{Frankfurt,
  Strikman, Day, and Sargsyan}}]{frankfurt93}
\bibinfo{author}{\bibfnamefont{L.}~\bibnamefont{Frankfurt}},
  \bibinfo{author}{\bibfnamefont{M.}~\bibnamefont{Strikman}},
  \bibinfo{author}{\bibfnamefont{D.}~\bibnamefont{Day}}, \bibnamefont{and}
  \bibinfo{author}{\bibfnamefont{M.}~\bibnamefont{Sargsyan}},
  \bibinfo{journal}{Phys. Rev. C} \textbf{\bibinfo{volume}{48}},
  \bibinfo{pages}{2451} (\bibinfo{year}{1993}).

\bibitem[{\citenamefont{Niyazov et~al.}(2004)}]{niyazov03}
\bibinfo{author}{\bibfnamefont{R.}~\bibnamefont{Niyazov}} \bibnamefont{et~al.}
  (\bibinfo{collaboration}{CLAS Collaboration}), \bibinfo{journal}{Phys. Rev.
  Lett.} \textbf{\bibinfo{volume}{92}}, \bibinfo{pages}{052303}
  (\bibinfo{year}{2004}).

\bibitem[{\citenamefont{Baghdasaryan et~al.}(2010)}]{bagh10}
\bibinfo{author}{\bibfnamefont{H.}~\bibnamefont{Baghdasaryan}}
  \bibnamefont{et~al.} (\bibinfo{collaboration}{CLAS Collaboration}),
  \bibinfo{journal}{Phys. Rev. Lett.} \textbf{\bibinfo{volume}{105}},
  \bibinfo{pages}{222501} (\bibinfo{year}{2010}).

\bibitem[{\citenamefont{Tang et~al.}(2003)}]{tang03}
\bibinfo{author}{\bibfnamefont{A.}~\bibnamefont{Tang}} \bibnamefont{et~al.},
  \bibinfo{journal}{Phys. Rev. Lett.} \textbf{\bibinfo{volume}{90}},
  \bibinfo{pages}{042301} (\bibinfo{year}{2003}).

\bibitem[{\citenamefont{Piasetzky et~al.}(2006)\citenamefont{Piasetzky,
  Sargsian, Frankfurt, Strikman, and Watson}}]{piasetzky06}
\bibinfo{author}{\bibfnamefont{E.}~\bibnamefont{Piasetzky}},
  \bibinfo{author}{\bibfnamefont{M.}~\bibnamefont{Sargsian}},
  \bibinfo{author}{\bibfnamefont{L.}~\bibnamefont{Frankfurt}},
  \bibinfo{author}{\bibfnamefont{M.}~\bibnamefont{Strikman}}, \bibnamefont{and}
  \bibinfo{author}{\bibfnamefont{J.~W.} \bibnamefont{Watson}},
  \bibinfo{journal}{Phys. Rev. Lett.} \textbf{\bibinfo{volume}{97}},
  \bibinfo{pages}{162504} (\bibinfo{year}{2006}).


\bibitem[{\citenamefont{Shneor et~al.}(2007)}]{shneor07}
\bibinfo{author}{\bibfnamefont{R.}~\bibnamefont{Shneor}} \bibnamefont{et~al.},
  \bibinfo{journal}{Phys. Rev. Lett.} \textbf{\bibinfo{volume}{99}},
  \bibinfo{eid}{072501} (\bibinfo{year}{2007}).


\bibitem[{\citenamefont{Subedi et~al.}(2008)}]{subedi08}
\bibinfo{author}{\bibfnamefont{R.}~\bibnamefont{Subedi}} \bibnamefont{et~al.},
  \bibinfo{journal}{Science} \textbf{\bibinfo{volume}{320}},
  \bibinfo{pages}{1476} (\bibinfo{year}{2008}).


\bibitem[{\citenamefont{Melnitchouk et~al.}(1997)\citenamefont{Melnitchouk,
  Sargsian, and Strikman}}]{meln97}
\bibinfo{author}{\bibfnamefont{W.}~\bibnamefont{Melnitchouk}},
  \bibinfo{author}{\bibfnamefont{M.}~\bibnamefont{Sargsian}}, \bibnamefont{and}
  \bibinfo{author}{\bibfnamefont{M.}~\bibnamefont{Strikman}},
  \bibinfo{journal}{Z. Phys. A} \textbf{\bibinfo{volume}{359}},
  \bibinfo{pages}{99} (\bibinfo{year}{1997}).

\bibitem[{\citenamefont{Klimenko et~al.}(2006)}]{klim06}
\bibinfo{author}{\bibfnamefont{A.~V.} \bibnamefont{Klimenko}}
  \bibnamefont{et~al.} (\bibinfo{collaboration}{CLAS Collaboration}),
  \bibinfo{journal}{Phys. Rev. C} \textbf{\bibinfo{volume}{73}},
  \bibinfo{pages}{035212} (\bibinfo{year}{2006}).

\bibitem[{\citenamefont{Mecking et~al.}(2003)}]{clas}
\bibinfo{author}{\bibfnamefont{B.A.}~\bibnamefont{Mecking}} \bibnamefont{et~al.},
  \bibinfo{journal}{Nucl. Inst. and Meth.} \textbf{\bibinfo{volume}{A503}},
  \bibinfo{pages}{513} (\bibinfo{year}{2003}).

\bibitem[{\citenamefont{Arrington}(2003)}]{arrington03}
\bibinfo{author}{\bibfnamefont{J.}~\bibnamefont{Arrington}},
  \bibinfo{journal}{Phys. Rev. C} \textbf{\bibinfo{volume}{68}},
  \bibinfo{pages}{034325} (\bibinfo{year}{2003}).

\bibitem[{\citenamefont{Baghdasaryan}(2007)}]{HBPhD}
\bibinfo{author}{\bibfnamefont{H.}~\bibnamefont{Baghdasaryan}}, Ph.D. thesis,
  \bibinfo{school}{Old Dominion University, Norfolk, VA}
  (\bibinfo{year}{2007}).

\bibitem[{\citenamefont{{CLAS GEANT Simulation}}(1997)}]{gsim}
\bibinfo{author}{\bibnamefont{{CLAS GEANT Simulation}}}, \bibinfo{type}{Tech.
  Rep.} (\bibinfo{year}{1997}),
  \urlprefix\url{http://nuclear.unh.edu/~maurik/Gsim/}.

\bibitem[{\citenamefont{Laget}(1987)}]{laget87}
\bibinfo{author}{\bibfnamefont{J.-M.} \bibnamefont{Laget}},
  \bibinfo{journal}{Phys. Rev. C} \textbf{\bibinfo{volume}{35}},
  \bibinfo{pages}{832} (\bibinfo{year}{1987}).

\bibitem[{\citenamefont{Laget}(1988)}]{laget88}
\bibinfo{author}{\bibfnamefont{J.-M.} \bibnamefont{Laget}},
  \bibinfo{journal}{J. Phys. G} \textbf{\bibinfo{volume}{14}},
  \bibinfo{pages}{1445} (\bibinfo{year}{1988}).

\bibitem[{\citenamefont{Laget}(1994)}]{laget94}
\bibinfo{author}{\bibfnamefont{J.-M.} \bibnamefont{Laget}},
  \bibinfo{journal}{Nucl. Phys.} \textbf{\bibinfo{volume}{A579}},
  \bibinfo{pages}{333} (\bibinfo{year}{1994}).

\bibitem[{\citenamefont{Laget}(2005)}]{laget05}
\bibinfo{author}{\bibfnamefont{J.-M.} \bibnamefont{Laget}},
  \bibinfo{journal}{Phys. Lett.} \textbf{\bibinfo{volume}{B609}},
  \bibinfo{pages}{49} (\bibinfo{year}{2005}).

\bibitem[{\citenamefont{Hajduk and Sauer}(1979)}]{hajduk79}
\bibinfo{author}{\bibfnamefont{C.}~\bibnamefont{Hajduk}} \bibnamefont{and}
  \bibinfo{author}{\bibfnamefont{P.~U.} \bibnamefont{Sauer}},
  \bibinfo{journal}{Nucl. Phys.} \textbf{\bibinfo{volume}{A322}},
  \bibinfo{pages}{329} (\bibinfo{year}{1979}).

\bibitem[{\citenamefont{Lacombe et~al.}(1981)}]{lac81}
\bibinfo{author}{\bibfnamefont{M.}~\bibnamefont{Lacombe}} \bibnamefont{et~al.},
  \bibinfo{journal}{Phys. Lett. B} \textbf{\bibinfo{volume}{101}},
  \bibinfo{pages}{139} (\bibinfo{year}{1981}).

\bibitem[{\citenamefont{Rvachev et~al.}(2005)}]{rvachev05}
\bibinfo{author}{\bibfnamefont{M.~M.} \bibnamefont{Rvachev}}
  \bibnamefont{et~al.} (\bibinfo{collaboration}{Jefferson Lab Hall A
  Collaboration}), \bibinfo{journal}{Phys. Rev. Lett.}
  \textbf{\bibinfo{volume}{94}}, \bibinfo{eid}{192302}
  (\bibinfo{year}{2005}).

\bibitem[{\citenamefont{Benmokhtar et~al.}(2005)}]{benmokhtar05}
\bibinfo{author}{\bibfnamefont{F.}~\bibnamefont{Benmokhtar}}
  \bibnamefont{et~al.} (\bibinfo{collaboration}{Jefferson Lab Hall A
  Collaboration}), \bibinfo{journal}{Phys. Rev. Lett.}
  \textbf{\bibinfo{volume}{94}}, \bibinfo{eid}{082305}
  (\bibinfo{year}{2005}).

\bibitem[{\citenamefont{Egiyan et~al.}(2007)}]{egi07}
\bibinfo{author}{\bibfnamefont{K.~S.} \bibnamefont{Egiyan}}
  \bibnamefont{et~al.} (\bibinfo{collaboration}{CLAS Collaboration}),
  \bibinfo{journal}{Phys. Rev. Lett.} \textbf{\bibinfo{volume}{98}},
  \bibinfo{pages}{262502} (\bibinfo{year}{2007}).

\bibitem[{\citenamefont{Piasetzky et~al.}(2007)}]{src07}
\bibinfo{author}{\bibfnamefont{E.}~\bibnamefont{Piasetzky}}
  \bibnamefont{et~al.}, \emph{\bibinfo{title}{{Jefferson Lab Expt E07-006,
  Studying Short-Range Correlations in Nuclei at the Repulsive Core Limit via
  the Triple Coincidence $(e,e'pN)$ Reaction}}} (\bibinfo{year}{2007}),
  \urlprefix\url{http://hallaweb.jlab.org/experiment/E07-006/}.

\end{thebibliography}
\end{document}